\documentclass[aps,prl,twocolumn,amsmath]{revtex4}
\usepackage{graphicx}
\begin{document}

\title{Disorder-induced rounding of certain quantum phase transitions}

\author{Thomas Vojta}
\affiliation{Department of Physics, University of Missouri-Rolla, Rolla, MO
65409}

\date{\today}

\begin{abstract}
We study the influence of quenched disorder on quantum phase transitions in
systems with over-damped dynamics. For Ising order parameter symmetry disorder
destroys the sharp phase transition by rounding because a static order
parameter can develop on rare spatial regions. This leads to an exponential
dependence of the order parameter on the coupling constant. At finite
temperatures the static order on the rare regions is destroyed. This restores
the phase transition and leads to a double-exponential relation between
critical temperature and coupling strength. We discuss the behavior based on
Lifshitz-tail arguments and illustrate the results by simulations of a model
system.
\end{abstract}



\maketitle

The influence of quenched disorder on phase transitions is an important problem
in condensed matter physics. Initially it was suspected that disorder destroys
any critical point \cite{Grinstein}. However, it was soon found that classical
continuous phase transitions generically remain sharp in the presence of weak
disorder.  If a clean critical fixed point (FP) fulfills the Harris criterion
\cite{Harris74} $\nu\ge 2/d$, where $\nu$ is the correlation length exponent
and $d$ is the spatial dimensionality, weak disorder is renormalization group
irrelevant. The system becomes asymptotically homogeneous at large length
scales. Even if the Harris criterion is violated, the transition will
generically be sharp. In this case the inhomogeneities either remain finite at
all length scales, leading to a finite disorder FP, or the relative magnitude
of the disorder diverges under coarse graining corresponding to an
infinite-randomness FP. A prominent example of the latter occurs in the
McCoy-Wu model \cite{McCoyWu}, a disordered 2D Ising-model in which the
disorder is perfectly correlated in one dimension. These correlations increase
the effects of the disorder.

An important aspect of phase transitions in disordered systems are the
Griffiths phenomena \cite{Griffiths}. They are caused by large spatial regions
that are devoid of any impurities and can be locally in the ordered phase even
if the bulk system is in the disordered phase. Since these regions are of
finite size, no true static order develops. The fluctuations of these regions
are very slow because they require changing the order parameter in a large
volume. Griffiths \cite{Griffiths} showed that this leads to a singular free
energy. In generic classical systems this is a weak effect, since the
singularity is only an essential one. An exception is the McCoy-Wu model
\cite{McCoyWu}. Here, the disorder correlations lead to stronger effects, with
the average susceptibility diverging in a finite region of the phase diagram.

Recently disorder effects have gained a lot of attention in the context of
quantum phase transitions \cite{QPT}. At these transitions order-parameter
fluctuations in space {\em and} time have to be considered. Quenched disorder
is time-independent, it is thus always correlated in one of the relevant
dimensions making disorder effects at quantum phase transitions generically
stronger than at classical transitions. A prototypical model is the random
quantum Ising ferromagnet. Its quantum phase transition in 1D
\cite{McCoy69,dsf9295,YoungRieger96} and 2D \cite{Pich98,Motrunich00} is
controlled by an infinite-randomness FP with activated rather than power-law
dynamical scaling. The Griffiths singularities are enhanced, too: Several
observables including the average susceptibility display power-law
singularities with continuously varying exponents over a finite region of the
disordered phase. Similar phenomena have also been found in quantum Ising spin
glasses \cite{ThillHuse95,gbh96,RiegerYoung96}.

The systems in which infinite-randomness FPs and quantum Griffiths phenomena
have been shown unambiguously all have undamped dynamics (a dynamical exponent
$z=1$ in the corresponding clean system). However, in itinerant electronic
systems the order parameter couples to fermionic modes leading to overdamped
dynamics (clean dynamical exponent $z>1$). A prototype example is the itinerant
quantum antiferromagnetic phase transition. The conventional perturbative
renormalization group for this transition \cite{BelitzKirkpatrick96} yields a
finite-disorder FP. However, by taking into account the effects of rare regions
it was shown \cite{NVBK99} that this FP is unstable, and the renormalization
group flow is towards large disorder. The meaning of this runaway flow is
presently not fully understood. In addition to the transition itself, quantum
Griffiths phenomena in itinerant systems have also attracted much attention
since they are of potential importance for a variety of heavy-fermion systems.
It has been suggested \cite{CastroNetoJones} that overdamped systems show
quantum Griffiths phenomena very similar to that of undamped systems. However,
recently it has been argued \cite{MillisMorrSchmalian} that for Ising symmetry
the overdamping prevents the rare regions from tunneling leading to
superpara\-magnetic rather than quantum Griffiths behavior.

In this Letter we reconsider the important question of disorder and rare
regions at quantum phase transitions. We show that for Ising order parameter
symmetry and Landau overdamped dynamics, the sharp quantum phase transition is
destroyed by rounding, because static order can develop on isolated rare
regions. Our results can be summarized as follows: The relation between the
order parameter $m$ and the distance $t$ from the clean critical point is
exponential for small $m$. The precise form depends on the disorder
distribution. For a Gaussian coupling constant distribution, $m$ is finite for
all $t$. Asymptotically it behaves as
\begin{subequations}
\begin{equation}
\log (m) \sim - t^{2-d/\phi}    \qquad \textrm{for } t\to\infty~.
\end{equation}
Here $\phi$ is the finite-size scaling shift exponent of the clean system. If
the disorder is of dilution type a finite order parameter starts to develop at
the clean transition, i.e., for $t<0$ \cite{huse_private}. For small $|t|$ it
behaves as
\begin{equation}
\log (m) \sim - |t|^{-d/\phi}  \qquad \textrm{for } t\to 0-~.
\label{eq:m-dilute}
\end{equation}
\label{eqs:m}
\end{subequations}
At finite temperatures the static order on the rare regions is destroyed, and a
finite interaction between them is required for long-range order. This restores
a sharp phase transition. The dependence of the critical temperature on $t$ is
double-exponential. For small $T_c$ we obtain
\begin{subequations}
\begin{eqnarray}
\log( -a \log T_c) &\sim& t^{2-d/\phi} \qquad \textrm{Gaussian},\\ \log( -a
\log T_c) &\sim& |t|^{-d/\phi} \qquad \textrm{dilution},
\end{eqnarray}
\label{eqs:tc}
\end{subequations}
Moreover, in finite-size samples strong sample-to-sample fluctuations occur if
the number of rare regions displaying static order becomes of order one.
Asymptotically for large linear system size $L$, the coupling strength $t_L$ at
which these fluctuations start behaves like
\begin{subequations}
\begin{eqnarray}
t_L &\sim& (\log L)^{1/(2-d/\phi)} \qquad \textrm{Gaussian},\\ |t_L| &\sim&
(\log L)^{-\phi/d} \qquad\qquad \textrm{dilution}. \label{eq:tL-dilute}
\end{eqnarray}
\label{eqs:tL}
\end{subequations}
Thus, finite size effects are suppressed only logarithmically. In the remainder
of this Letter, we sketch the derivation of these results, illustrate them by
numerical results from a model system and discuss their relevance for
experiments as well as simulations.

For definiteness we consider the antiferromagnetic quantum phase transition of
itinerant electrons with Ising spin symmetry. The Landau-Ginzburg-Wilson free
energy functional of the clean transition reads
\cite{BelitzKirkpatrick96,Hertz76}
\begin{equation}
S = \int dx\,dy\ m(x)\,\Gamma(x,y)\,m(y)
    + u\int dx\ m^4(x)
\label{eq:action}
\end{equation}
Here $x\equiv ({\bf x},\tau)$ comprises position ${\bf x}$ and imaginary time
$\tau$, and $\int dx \equiv \int d{\bf x}\int_{0}^{1/T}d\tau$. $\Gamma(x,y)$ is
the bare two-point vertex, whose Fourier transform is $\Gamma({\bf q},\omega_n)
= (t + {\bf q}^2 + \vert\omega_n\vert)$. Here $t=(g-g_c)/g_c$ is the distance
of the coupling constant $g$ from the critical point. The dynamical part of
$\Gamma$ is proportional to $|\omega_n|$ reflecting the overdamping of the
dynamics (undamped dynamics leads to $\omega_n^2$).
Quenched disorder is introduced by making $t$ a random function of position, $t
\to t + \delta t({\bf x})$. We consider two different disorder distributions.
The first is a Gaussian distribution with zero mean and a correlation function
$\langle \delta t({\bf x})\delta t({\bf y})\rangle = \Delta^2 \delta({\bf
x}-{\bf y})$. The second disorder distribution is of dilution type, $\delta
t({\bf x})=0$ everywhere except on randomly distributed finite-size islands
(impurities) of spatial density $c$ where $\delta t({\bf x})=W>0$.

In the presence of disorder there are rare large spatial regions which are
locally in the ordered phase while the bulk system is not. For the Gaussian
disorder distribution which is unbounded, these regions exist for all $t$. For
the dilution case they appear below the transition of the clean system, i.e.
for $t<0$. At zero temperature a single such region is equivalent to a
classical Ising model in a rod-like geometry. It is finite in the $d$ space
dimensions but infinite in imaginary time. If the interaction in the time
direction is short-ranged, as is the case for undamped dynamics, true static
order cannot develop on such a rare region. Instead, the order parameter
displays slow fluctuations leading to quantum Griffiths phenomena. This is
drastically different in a system with overdamped dynamics. The linear
frequency dependence in $\Gamma$ is equivalent to a long-range interaction in
time of the form $(\tau-\tau')^{-2}$. 1D Ising models with $1/r^2$ interaction
are known to develop long-range order at finite temperatures \cite{Ising1r2}.
Thus, in our quantum system, true static order develops on those rare regions
which are locally in the ordered phase. In agreement with Ref.\
\cite{MillisMorrSchmalian} we therefore do not find the usual quantum Griffiths
phenomena. Once static order has developed on a few isolated rare regions, an
infinitesimally small interaction or an infinitesimally small symmetry-breaking
field are sufficient to align them. Consequently, a macroscopic order parameter
arises.

We now use Lifshitz tail arguments \cite{Lifshitz} to derive the leading
thermodynamic behavior for small order parameter $m$. In the dilution case, the
probability $w$ to find a region of linear size $L_R$ devoid of any impurities
($\delta t =0$) is given by $w \sim \exp( -c L_R^d)$ (up to pre-exponential
factors). Such a rare region develops static order at a some $t_c(L_R) < 0$.
Finite size scaling yields $|t_c(L_R)| \sim L_R^{-\phi}$ where $\phi$ is the
finite-size scaling shift exponent of the clean system \cite{shiftexponent}.
Thus, the probability for finding a rare region which becomes critical at $t_c$
is given by
\begin{subequations}
\begin{equation}
w(t_c) \sim \exp  (-B ~|t_c|^{-d/\phi}) \qquad \textrm{for } t\to 0-~.
\end{equation}
For a Gaussian distribution similar arguments \cite{HalperinLax} give
\begin{equation}
w(t_c) \sim  \exp ( -\bar B ~t_c^{2-d/\phi} )    \qquad \textrm{for }
t\to\infty~.
\end{equation}
\label{eqs:wtc}
\end{subequations}
Here $B$ and $\bar B$ are constants. The total order parameter $m$ is obtained
by integrating over all rare regions which are ordered at $t$, i.e., all rare
regions having $t_c>t$. This leads to eqs. (\ref{eqs:m}). Note that the
functional dependence on $t$ of the order parameter on a given island is of
power-law type and thus only influences the pre-exponential factors.

At finite temperatures the static order on the rare regions is destroyed, and a
finite interaction of the order of the temperature is necessary to align them.
This means a sharp phase transition is recovered. To estimate the transition
temperature we note that the interaction between two rare regions depends
exponentially on their spatial distance $r$, $E_{int} \sim \exp(-r/\xi_0)$,
where $\xi_0$ is the bulk correlation length. The typical distance $r$ itself
depends exponentially on $t$ via $r \sim w^{-1/d}$. The leading dependence of
the critical temperature $T_c$ on $t$ is thus
\begin{equation}
T_c \sim \exp(-r/\xi_0) \sim \exp( -a~ w^{-1/d}/\xi_0)~,
\end{equation}
where $a$ is a constant. Inserting eqs.\ (\ref{eqs:wtc}) for $w$ leads to eqs.\
(\ref{eqs:tc}). Above this exponentially small critical temperature the rare
regions essentially behave classically.

We now turn to finite-size effects at zero temperature. Since the total order
parameter is the sum of contributions of many independent islands, finite
size-effects in a macroscopic sample are governed by the central limit theorem.
However, for $t\to\infty$ (Gaussian distribution) or $t\to 0-$ (dilution) very
large and thus very rare regions are responsible for the order parameter. The
number $N$ of rare regions which start to order at $t_c$ in a sample of size
$L$ behaves like $N \sim L^d ~w(t_c)$. When $N$ becomes of order one, strong
sample-to-sample fluctuations arise. Using eqs.\ (\ref{eqs:wtc}) for $w(t_c)$,
this leads to eqs.\ (\ref{eqs:tL}).

To illustrate the rounding of the phase transition we now present numerical
data for a model system, viz., a disordered classical Ising model with two
space dimensions and one time-like dimension. The disorder is of dilution type
and completely correlated in time direction. The interaction is short-ranged in
the space directions but infinite-ranged in the time-like direction. This
simplification retains the crucial property of static order on the rare regions
but allows us to treat system sizes large enough for the investigation of
exponentially rare events. The Hamiltonian of the model reads
\begin{equation}
H= - \frac 1 L_\tau\sum_{\langle{\bf x,y}\rangle,\tau,\tau'}   S_{{\bf x},\tau}
S_{{\bf y},\tau'}
   - \frac 1 L_\tau\sum_{{\bf x  },\tau,\tau'} J_{\bf x} S_{{\bf x},\tau} S_{{\bf x},\tau'}
\label{eq:toy}
\end{equation}
Here ${\bf x}$ and $\tau$ are the space and time coordinates, respectively.
$L_\tau$ is the system size in time direction and $\langle{\bf x,y}\rangle$
denotes pairs of nearest neighbors. $J_{\bf x}$ is a quenched binary random
variable with the distribution $P(J) = (1-c)~ \delta(J-1) + c~ \delta(J)$. In
this classical model $L_\tau$ takes the role of the inverse temperature in the
corresponding quantum system and the classical temperature takes the role of
the coupling constant $g$. Because the interaction is infinite-ranged in time,
the time-like dimension can be treated in mean-field theory. For
$L_\tau\to\infty$, this leads to a set of coupled mean-field equations (one for
each {\bf x})
\begin{equation}
 m_{\bf x} = \tanh \beta~ [ J_{\bf x} m_{\bf x} + {\sum_{{\bf y}({\bf x})}}' m_{\bf y} + h]~,
\label{eq:mf}
\end{equation}
where we have added a very small symmetry-breaking magnetic field, $h=10^{-8}$.
These equations are solved numerically in a self-consistency cycle.

\begin{figure}
\centerline{\includegraphics[width=7.3cm]{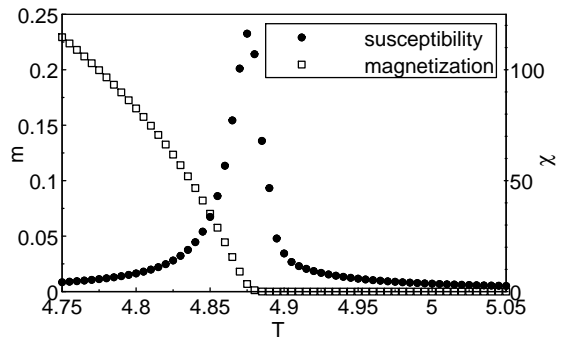}} \caption{Magnetization
and susceptibility ($L=100, ~c=0.2$).} \label{Fig:overview}
\end{figure}

\begin{figure}
\centerline{\includegraphics[width=7.3cm]{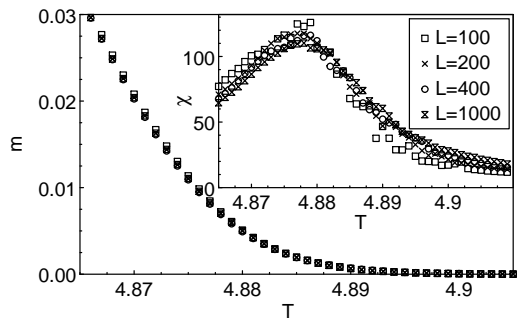}} \caption{Magnetization
and susceptibility close to the seeming transition for different
         system sizes.}
\label{Fig:rounding}
\end{figure}

Fig.\ \ref{Fig:overview} shows the total magnetization and the susceptibility
(which corresponds to the inverse energy gap of the quantum system) as
functions of temperature for linear size $L=100$ and impurity concentration
$c=0.2$. The data are averages over 200 disorder realizations \cite{average}.
At a first glance these data suggest a sharp phase transition close to
$T=4.88$. However, a closer investigation, Fig.\ \ref{Fig:rounding},
shows that the singularities are rounded. If this rounding was a conventional
finite-size effect the magnetization curve should become sharper with
increasing $L$ and the susceptibility peak should diverge. This is not the case
here, instead the transition remains rounded for $L\to\infty$.

\begin{figure}
\centerline{\includegraphics[width=7.3cm]{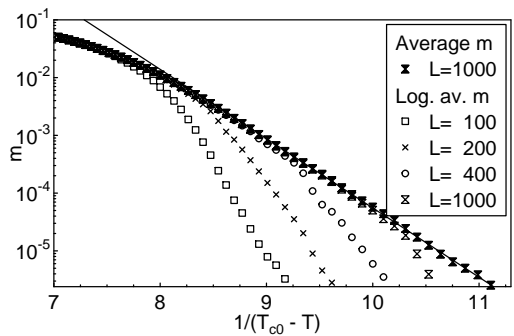}} \caption{Log($m$) as
a function of distance from the clean critical point.
         The solid line is a fit of the average magnetization to eq.\ (\ref{eq:m-dilute}) with $\phi=2$.
         The logarithmically averaged data show the onset of the sample-to-sample fluctuations. }
\label{Fig:logarithm}
\end{figure}

For comparison with the analytical results, Fig.\ \ref{Fig:logarithm} shows the
logarithm of the average magnetization as a function of $1/(T_{c}^0-T)$ where
$T_{c}^0=5$ is the critical temperature of the clean system ($c=0)$.
The data follow eq.\ (\ref{eq:m-dilute}) over almost four orders of magnitude
in $m$ with the expected shift exponent of $\phi=2$. This figure also shows
"typical", i.e., logarithmically averaged magnetization data for different
system sizes. Deviations between the typical and the average values (which are
essentially size-independent) reflect strong sample-to-sample fluctuations. The
data show that the onset $T_L$ of these fluctuations shifts to larger
temperatures with increasing system size. A more detailed analysis shows that
$t_L = (T_L-T_{c}^0)$ follows eq.\ (\ref{eq:tL-dilute}) in good approximation.

In the remaining paragraphs we discuss the generality of the results and their
consequences for experiments and simulations. The rounding of the transition is
due to the fact that at zero temperature static order can develop on a finite
spatial region in a Landau-damped system. We thus expect all quantum phase
transitions with discrete order parameter symmetry and overdamped dynamics
(i.e., a low frequency dependence $\sim |\omega|$ or slower in the Gaussian
propagator) to be rounded by disorder. Systems with continuous symmetry behave
differently. It is known \cite{bruno01} that classical 1D XY and Heisenberg
systems develop long-range order at finite $T$ only if the interaction falls
off more slowly than $1/r^2$. Consequently, in a quantum system at zero
temperature static order on a rare region only develops if the frequency
dependence in the Gaussian propagator is slower than $|\omega|$. The itinerant
ferromagnetic or antiferromagnetic quantum phase transitions with XY or
Heisenberg symmetries will thus remain sharp in the presence of disorder
\cite{AbanovChubukov}.

The rounding of the quantum phase transition leads to an unusual phenomenology
in experiments. Data taken at larger order parameter and not too low
temperature do not resolve the exponentially small order parameter tail but
probe the rounded transition. However, with increasing precision and decreasing
temperature the apparent critical point shifts further and further towards the
disordered phase, accompanied by the onset of strong sample-sample
fluctuations. Similar effects will also occur in simulations. In a recent
Monte-Carlo simulation of a related model \cite{us_mc} a sharp transition was
found controlled by an infinite randomness fixed point. However, in this
simulation the linear system sizes are below $L=30$, therefore it likely probes
the rounded transition rather than the exponentially small magnetization tail.

In this Letter we have concentrated on the behavior of the system for very
small order parameter values, i.e., in the "tail region" of the rounded
transition. The above arguments suggest that the properties of the rounded
transition itself are experimentally also very important because they control
the behavior in a wide pre-asymptotic region. However, these properties are
likely to be non-universal, so far they are not very well understood.

We thank D. Belitz, D.\ Huse, T.R.\ Kirkpatrick, J.\ Schmalian and R.\ Sknepnek
for helpful discussions, and we acknowledge support from the University of
Missouri Research Board. Part of this work has been performed at the Aspen
Center for Physics.



\begin{thebibliography}{99}
\bibitem{shiftexponent}The upper critical dimension of the clean itinerant antiferromagnetic quantum
     phase transition is $d_c^+=2$. Thus, hyperscaling is not valid in 3D and $\phi \ne 1/\nu$. For our
     purpose the exact value of $\phi$ is not important.
\bibitem{average}Thermodynamic quantities involve averaging over the whole system. Thus ensemble averages
     rather than typical values give the correct infinite system
     approximation.
\end{thebibliography}
\end{document}